\documentclass[prl,twocolumn,showpacs]{revtex4}
\usepackage{graphicx}

\begin{document}

\title{Nonequilibrium fluctuations in a resistor}
\date{\today}
\author{N. Garnier}
\email{nicolas.garnier@ens-lyon.fr}
\author{S. Ciliberto}
\affiliation{Laboratoire de Physique, CNRS UMR 5672, 
Ecole Normale Sup\'erieure de Lyon, 46 All\'ee d'Italie, 69364 Lyon CEDEX 07, France}

\begin{abstract}

In small systems where relevant energies are comparable to thermal
agitation, fluctuations are of the order of average values. In systems
in thermodynamical equilibrium, the variance of these fluctuations can
be related to the dissipation constant in the system, exploiting the
Fluctuation-Dissipation Theorem (FDT). In non-equilibrium steady
systems, Fluctuations Theorems (FT) additionally describe symmetry
properties of the probability density functions (PDFs) of the
fluctuations of injected and dissipated energies. We experimentally
probe a model system: an electrical dipole driven out of equilibrium by
a small constant current $I$, and show that FT are experimentally
accessible and valid. Furthermore, we stress that FT can be used to
measure the dissipated power $\bar{\cal P}=RI^2$ in the system by just
studying the PDFs symmetries.

\pacs{05.40.-a, 05.70.-a, 07.50.-e, 84.30.Bv}
\end{abstract}

\maketitle

{\bf Introduction}

In the last decade, Fluctuation Theorems (FT)~\cite{ECM,GC} appeared in
nonequilibrium statistical physics. These new theorems relate the
asymmetry of fluctuations of energies (or powers) with the dissipated
power required to maintain a nonequilibrium steady state of a system.
Thus FT give a measure of the distance from equilibrium. Indeed, FT can
be extremely useful to probe nonequilibrium states in nanophysics and
biophysics where energies involved are typicaly small and thermal
agitation cannot be neglected: in those systems, the variance of
fluctuations is of the order of the average values. Moreover, standard
Fluctuation-Dissipation Theorem (FDT)~\cite{FDT,Einstein,Smoluchowski}
is derived for equilibrium systems only and so it may fail to describe
states far from equilibrium. FT are a generalization of FDT out of
---~possibly far from~--- equilibrium. In order to safely apply FT in
complex systems ---~{\em e.g.} in biophysics or nanotechnologies~--- it
is important to test them in all simple cases where theoretical
predictions can be accurately checked. In spite of the large theoretical
effort, just a few experiments have been conducted on this
topic~\cite{Wang,Carberry}. For this reason, we test in this letter
their use on a simple electrical system.

Electrical systems are particuliarly interesting, because they are one
of the first where FTD was formalized: Johnson~\cite{Johnson} and
Nyquist~\cite{Nyquist} related equilibrium fluctuations of voltage $U$
across a dipole with the resistive part of this dipole. Moreover, all
parameters in the setup can easily be controled. Thus, out of
equilibrium, electrical circuits are a key example to stress the
differences between FT and FDT, and to verify the validity of FT.

\begin{figure}
\begin{center} \includegraphics[scale=0.6]{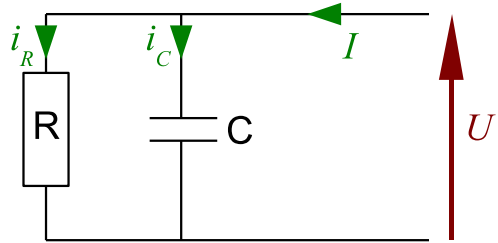}
\end{center}
\caption{
Model circuit : an electrical dipole is composed of a resistive part $R$
	and a capacitive part $C$. Due to thermal fluctuations of charges positions, a
	fluctuating voltage $U$ is observed. We drive the system away from equilibrium 
	by imposing a constant flux of electrons, via a constant current $I$.}
\label{fig:schema}
\end{figure}

Our system is an electrical dipole constituted of a resistance $R$ in
parallel with a capacitor $C$ (Fig.~\ref{fig:schema}). We drive it out
of equilibrium by making a constant current $I$ flow in it. Noting $k_B$
the Boltzmann constant and $T$ the absolute temperature, the injected
power is typically of some $k_{\rm B}T$ per second, of the same order as
in biophysics or nanoscale physics experiments. This represent the
fundamental case of a system in contact with two different (electrons-)
reservoirs, one of the simplest and most fundamental problems of
nonequilibrium physics~\cite{Derrida}. Using a powerful analogy with a
forced Langevin equation~\cite{Cohen1}, a precise formulation of FT was
recently given~\cite{Cohen} in this case. Nyquist FDT as well as FT are
checked experimentally, by looking at the injected power and the
dissipated heat in the system.

{\bf Experimental setup}

The circuit we use is composed of a resistor in parallel with a
capacitor, as depicted on Fig.1. The resistance is a standard metallic
one of nominal value $R=9.52$ M$\Omega$. In parallel, we have an
equivalent capacitor of value $C=280$ pF. This accounts for the
capacitance of the all set of coaxial connectors and cables that we
used. The time constant of the circuit is $\tau_0\equiv RC =2.67$ms.
Using a 50 G$\Omega$ resistance, we inject in the circuit a constant
current $I$ ranging from 0 to $6 \times 10^{-13}$ A. This current
corresponds to an injected power $I^2 R$ ranging from 0 to $1000 k_{\rm
B}T$/s. Experiments are conducted at room temperature $T=300$ K.
The typical values of the injected energy for, e.g., $I=1.4 \times
10^{-13}$ A and $\tau=10\tau_0$ are of order of a few hundreds of
$k_{\rm B} T$, which is small enough to ensure that the resistance is
not heating; expected changes of temperature are estimated to be less
than $10^{-14}$ K over a one hour experiment. Moreover, the resistance
is thermostated: all the heat dissipated by Joule effect is absorbed by
the thermal bath.

The fluctuating voltage $U$ across the dipole is measured with a
resolution of $10^{-11}$ V sampled at 819.2 Hz. This is achieved by
first amplifying the signal by $10^4$, using a FET amplifier, with a 4
G$\Omega$ input impedance, a voltage noise level of 5 nV/$\sqrt{\rm Hz}$
and a current noise of $10^{-15}$ A/$\sqrt{\rm Hz}$. The signal is then
digitized with a 24-bits data acquisition card at frequency 8192 Hz,
and decimated by averaging ten consecutive points.

{\bf Fluctuation-Dissipation theorem}

The electrical dipole of Fig.~\ref{fig:schema} is a pure resistance $R$
in parallel with a capacitor $C$; its complex impedance reads
$Z(f)=1/(1/R+i2\pi RCf)$ where $i^2=-1$ and $f$ is the frequency. The
effective dissipative part is the real part $\Re(Z)$ of $Z$. It was
experimentally observed by Johnson~\cite{Johnson}, and then demonstrated
by Nyquist~\cite{Nyquist} that the potential difference $U$ across the
dipole fluctuates with a stationary power spectral density $S(f)df$ such
that
\begin{equation}
	 S(f) df = 4 k_{\rm B} T \Re(Z) df \,,
\label{eq:N1}
\end{equation}
At equilibrium $I=0$, in average no current is flowing in the circuit, and 
mean $U$ is zero. Integrating over all positive frequencies, one
gets the variance of $U$:
\begin{equation}
	 \langle U^2 \rangle = \frac{k_{\rm B} T R}{\tau_0} = \frac{k_{\rm B}T}{C}  \,.
\label{eq:N2}
\end{equation}
Eqs. (\ref{eq:N1}) and (\ref{eq:N2}) are the expressions of the
Fluctuation-Dissipation theorem (FDT) for electrical circuits.

\begin{figure}
\begin{center}
\includegraphics[width=8cm]{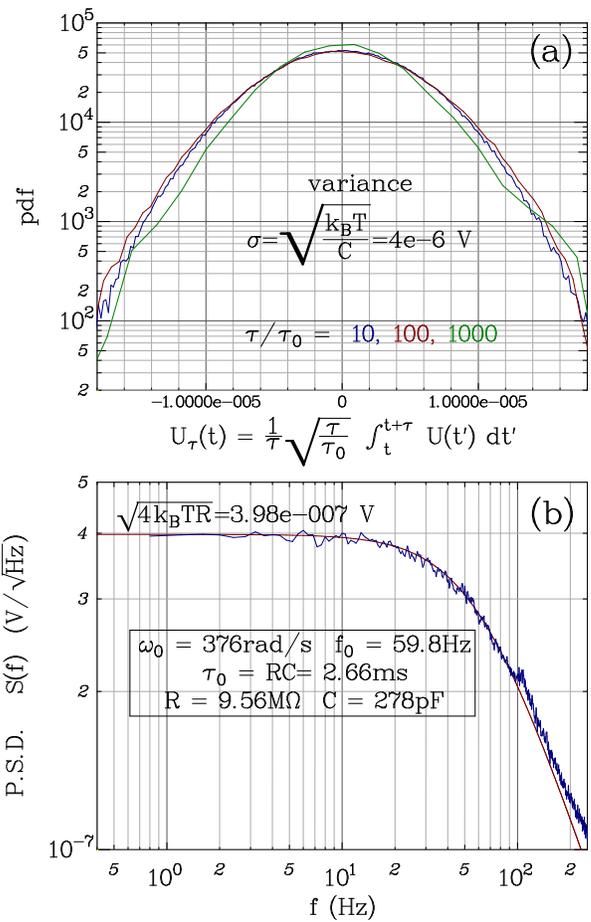}
\end{center}
\caption{ (a): For $I=0$ A,
Johnson-Nyquist noise has a Gaussian distribution. 
Relation (\ref{eq:N2}) is verified. 
(b): The noise is white up to the cutoff frequency $f_0$ of 
the $RC$ dipole. We use FDT to extract from the energy of the noise
the value of the resistive, dissipative part of the circuit $R$: for low frequency, the noise level
spectral density is constant, equal to $\sqrt{4 k_B T R}$ in a band $[f;~f+{\rm d}\!f]$.
A Lorentzian fit of the spectrum (thin line) additionally gives $\tau_0=RC$. 
}
\label{fig:FDT}
\end{figure}

The exact value of our capacitance was determined by fitting the power
density spectrum of equilibrium fluctuations (at imposed $I=0$ A) by a
Lorentzian low-pass transfer function (eq.(\ref{eq:N1})), as illustrated
on Fig.~\ref{fig:FDT}b. Application of FDT leads with a very good
accuracy to the determination of $R$, in perfect accordance with the
measured nominal value (Fig.2). When $I=1.4 \times 10^{-13}$ A, we found
the same power spectral density for $U$, and performing the same
treatments gave the same estimates of $R$ and $C$. We therefore conclude
that FDT is still holding in our system driven out of equilibrium.

{\bf Fluctuation theorems}

The power injected in the circuit is ${\cal P}_{\rm in}=U I$, but only 
the resistive part dissipates, so the dissipated power is
${\cal P}_{\rm diss}=U i_{\rm R}$, where $i_{\rm R}$ is the current
flowing in the resistor (Fig.~\ref{fig:schema}). As already
noted~\cite{Aumaitre}, in average, one expects $\langle {\cal P}_{\rm
in} \rangle = \langle {\cal P}_{\rm diss} \rangle \equiv \bar{\cal P}$,
where the brackets stand for time average over sufficiently long times
compared to $\tau_0$. This is very well checked in our experiment.
${\cal P}_{\rm in}$ and ${\cal P}_{\rm diss}$ fluctuate in time because
$U$ itself is fluctuating. If one assumes that fluctuations of $U$ have
a Gaussian distribution, which is the case at equilibrium when $I=0$,
then ${\cal P}_{\rm in}$ has also a Gaussian distribution, because $I$ is
constant. On the contrary, $i_{R}$ fluctuates, as we see from Kirchoff's
laws: 
\begin{equation} 
I = i_R + C \frac{{\rm d}U} {{\rm d}t} \,, \qquad {\rm so} \qquad 
{\cal P}_{\rm in} = {\cal P}_{\rm diss} + \frac{1}{2} C \frac{{\rm d}U^2} {{\rm d}t}
\,, 
\end{equation}

and therefore, the probability distribution of ${\cal P}_{\rm diss}$ is
not Gaussian~\cite{Cohen}. It is worth noting that for large current
$I$, some orders of magnitude larger than the one we use, ${\cal P}_{\rm
in}$ and ${\cal P}_{\rm diss}$ will be much larger than the conservative
part $\frac{C}{2} \frac{{\rm d}U^2} {{\rm d}t}$ and therefore the
probability distributions of both the injected and dissipated power will
be practically equal and usually Gaussian, as it is expected in
macroscopic systems.

We call $\langle g \rangle_\tau(t) = \frac{1}{\tau} \int_{t}^{t+\tau}
g(t') dt'$ the time-averaged value of a function $g$ over a time $\tau$.

Reasoning with energies instead of powers, we define $W_\tau(t) = \tau
\langle {\cal P}_{\rm in} \rangle_{\tau}(t)$, the energy injected in the
circuit during time $\tau$, analogous to the work performed on the
system (positive when received by the system). In the same way, we write
$Q_\tau(t) = \tau \langle {\cal P}_{\rm diss} \rangle_{\tau}(t)$, the
energy dissipated by Joule effect during time $\tau$, analogous to the
heat dissipated by the system (positive when given by the system to the
thermostat). We used values of $\tau$ spanning from fractions of
$\tau_0$ up to hundreds of $\tau_0$.

\begin{figure}
\begin{center}
\includegraphics[width=8cm]{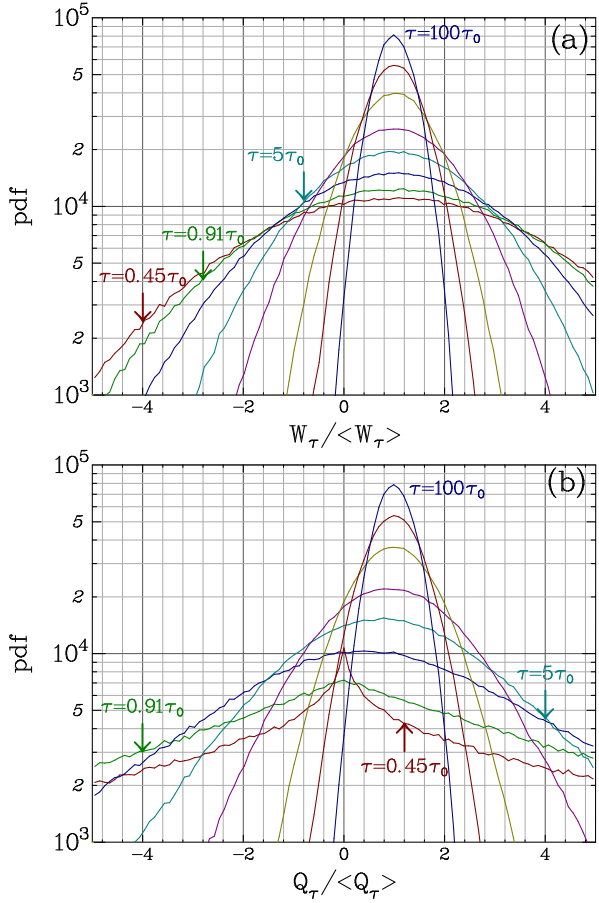}
\end{center}
\caption{
Histograms of $W_\tau$ and $Q_\tau$ when a current $I=1.4 \times 10^{-13}$A is flowing in the dipole.
This corresponds to $\langle W_\tau \rangle = \langle Q_\tau \rangle 
= \tau \bar{\cal P}$ with $\bar{\cal P} = RI^2=45 k_{\rm B} T$/s.
}
\label{fig:histos}
\end{figure}

\begin{figure}
\begin{center}
\includegraphics[width=8cm]{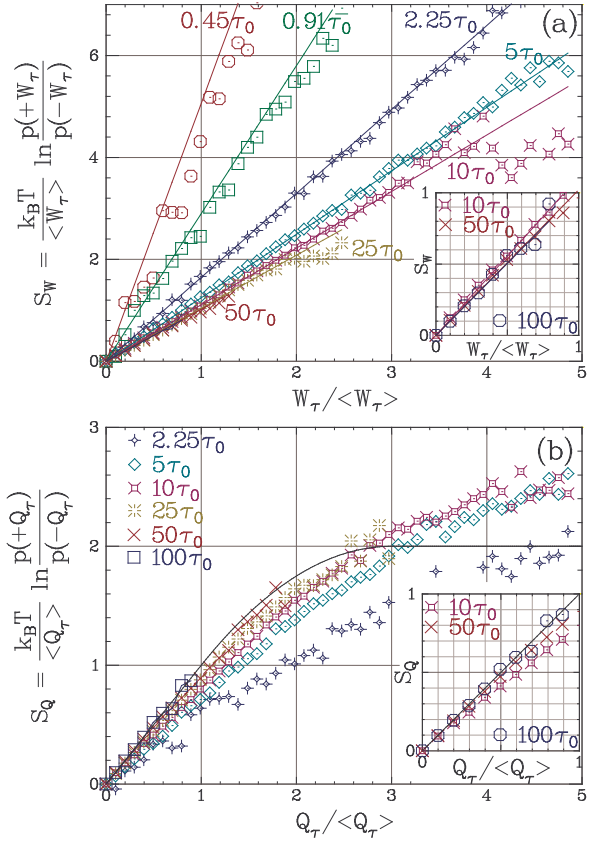}
\end{center}
\caption{ Normalized
symmetry functions $S_W$ and $S_Q$ for $W_\tau$ and $Q_\tau$ when $I=1.4 \times 10^{-13}$A 
(same conditions as Fig.~\ref{fig:histos}).  
(a): For any $\tau$, $S_W$ is a linear function of $a=W_\tau/\langle W_\tau \rangle$. 
Straight lines are theoretical predictions from~\cite{Cohen}.
For $\tau \rightarrow \infty$, $S_W(a)$ tends to have a slope 1.
(b): On the contrary, $S_Q(a)$ tends to a limit function (continuous black curve) 
which is a straight line of slope 1 for $a<1$ only.
}
\label{fig:sym_func}
\end{figure}

For a given value of $I$, we measure $U(t)$ and compute ${\cal P}_{\rm
in}$ and ${\cal P}_{\rm diss}$. We then build the probability density
functions of cumulated variables $W_\tau$ and $Q_\tau$ using $10^6$
points; their typical distributions are plotted on
Fig.~\ref{fig:histos}. As expected, fluctuations of $W_\tau$ are
Gaussian for any $\tau$ whereas heat fluctuations are not for small
values of $\tau$. They are exponential for small $\tau$: large
fluctuations of heat $Q_\tau$ are more likely to occur than large
fluctuations of work $W_\tau$.

Then we study the normalized symmetry function:
\begin{equation}
S_E(\tau,a) \equiv \displaystyle \frac{k_{\rm B} T}{\tau \bar{\cal P}} \ln \frac{p(E_\tau=a)}{p(E_\tau=-a)} \,,
\label{eq:FT}
\end{equation}
where $E_\tau$ stands for either $W_\tau/\langle W_\tau \rangle$ or
$Q_\tau/\langle Q_\tau\rangle$. If the Fluctuation Theorem for the work
$W_\tau$ holds, then one should have~\cite{Cohen1, Cohen}, for large
enough $\tau$, the relationship $\lim_{\tau \rightarrow \infty}
f_W(\tau,a)=a$. In contrast, if the Fluctuation Theorem for the heat
$Q_\tau$ holds, then for $\tau \rightarrow \infty$, the asymptotic
values of $S_Q(\tau,a)$ are $S_Q^\infty(a)=a$ for $a\le 1$,
$S_Q^\infty(a)=2$ for $a\ge 3$, and there is a continuous parabolic
connection for $1\le a \le 3$ that has a continuous
derivative~\cite{Cohen1,Cohen}. From histograms of
Fig.~\ref{fig:histos}, we compute the symmetry functions $S_W(\tau,a)$
and $S_Q(\tau,a)$ (Fig.~\ref{fig:sym_func}).

\paragraph{Work fluctuations}

\begin{figure}
\begin{center}
\includegraphics[width=8cm]{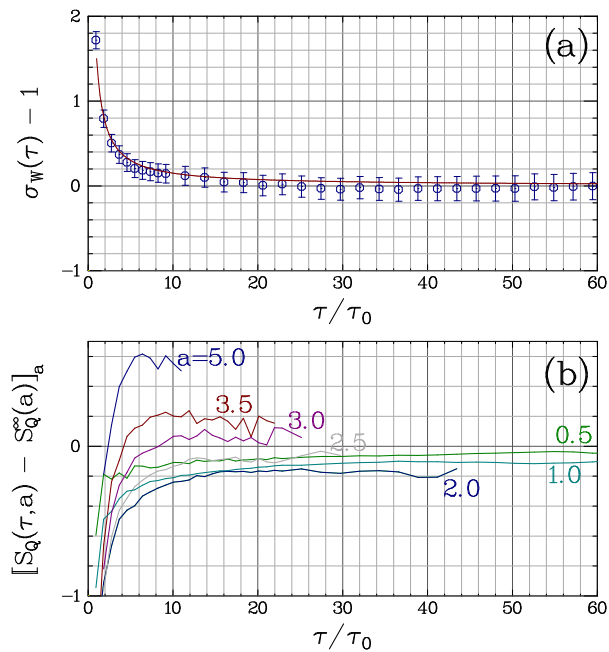}
\end{center}
\caption{
(a): dependence on $\tau$ of the slope of $S_W$ in Fig.~\ref{fig:sym_func}.
Slope is converging from above.
(b): distance between $S_Q(\tau)$ for finite $\tau$ and theoretical prediction 
of $S_Q$ for infinite $\tau$, for several values of $a=Q/\langle Q \rangle$.
for $a<3$, convergence is from below, whereas it is from above for $a>3$.
}
\label{fig:convergence}
\end{figure}

First, for any given $\tau$ we checked that the symmetry function
$S_W(\tau,a)$ is linear in $a$ (Fig.~\ref{fig:sym_func}a). We measured
the corresponding proportionality coefficient $\sigma_W(\tau)$ such that
$S_W(\tau,a)=\sigma_W(\tau)a$. This coefficient $\sigma_W(\tau)$ tends
to 1 when $\tau$ is increased (see Fig.~\ref{fig:convergence}). 


\paragraph{Heat fluctuations}

We found that $S_Q(\tau,a)$ is linear in $a$ only for $a<1$, as
expected~\cite{Cohen1,Cohen}. Again, as $\tau \rightarrow \infty$, the
limit slope of the symmetry function is 1 whereas for $a>3$,
$S_Q(\tau,a)$ tends to two.

\paragraph{Asymptotic symmetry functions and convergence}

In~\cite{Cohen,Cohen1}, expressions for the convergence towards the
asymptotic limits $S_W^\infty(a)$ and $S_Q^\infty(a)$ are given in terms
of $\tau$. We can check these predictions with our data. The 
convergence for the work $W_\tau$ is very well reproduced by these
predictions, as can be seen on Fig.~\ref{fig:sym_func}a: continuous
straight lines are theoretical predictions for small values of $\tau$,
using eqs. (9) and (10) from~\cite{Cohen}, with no adjustable
parameters. On Fig.~\ref{fig:convergence}a the slope $\sigma_W(\tau)$ of
the experimental symmetry function is plotted. The continuous line is the
prediction of~\cite{Cohen}, which perfectly agrees with our data.

For the heat, we distinguish two regimes for the convergence towards the
asymptotic symmetry function. For $a=Q_\tau/\langle Q \rangle <3$, we
find that when $\tau$ is increased, symmetry functions are converging to
the asymptotic function from below (Fig.~\ref{fig:sym_func}), which is
the opposite of what is observed for the work. On the contrary, for
$a>3$, convergence to the asymptotic function is from above, thus
enhancing the peculiarity of the point $a=3$. On
Fig.~\ref{fig:convergence}, we have plotted the evolution of
$S_Q(\tau,a)$ versus $\tau$ for several fixed values of $a$. The
convergence from above for $a\ge 3$ and from below for $a<3$ is clear.
For increasing values of $\tau$, only smaller and smaller values of $a$
are accessible because of the averaging process. Therefore the
accessible values of $a$ are quickly lower than 1, and only the linear
part of $S_Q(\tau,a)$ can be experimentally tested. Nevertheless, for
intermediate time scales $\tau$ we see in Fig.~\ref{fig:sym_func} that
the data converge towards the theoretical asymptotic nonlinear symmetry
function~\cite{Cohen,Cohen1} (smooth curve in Fig.~\ref{fig:sym_func}b).

We observed that the convergence reproduced on
Fig.~\ref{fig:convergence} depends on the injected current $I$, as
pointed out in~\cite{Cohen}. Other experiments with a larger current
($\bar{\cal P}=186 k_{\rm B}T$/s) give a faster convergence;
corresponding results will be reported elsewhere.

{\bf Conclusions}

We have shown experimentally that the asymmetry of the probability
distribution functions of work and heat in a simple electrical circuit
driven out of equilibrium by a fixed current $I$, is linked to the
averaged dissipated power in the system. The recently proposed
Fluctuation Theorems for first order Langevin systems are then
experimentally confirmed. Exploiting formula~(\ref{eq:FT}), FT can be
used to measure an unknown averaged dissipated power $\bar{\cal P} =
\lim_{\tau \rightarrow \infty} \frac{Q_\tau}{\tau}$ by using only the
symmetries of the fluctuations, {\em i.e.} computing $S_W$ or $S_Q$ and
measuring their asymptotic slope. 

We operated with energies of order of $k_BT$ in order to have strong
fluctuations compared to the averaged values. It is worth noting that as
the driving current is increased up to macroscopic values, the
fluctuations become more and more negligible, therefore Fluctuation
Theorems become harder and harder to use, and therefore less relevant.


\end{document}